\documentclass[sort&compress,final]{aipproc}\layoutstyle{6x9}
\usepackage{amsmath}\usepackage{amssymb}\usepackage{booktabs}
\begin{document}\title{Local-Duality QCD Sum Rules for
Pseudoscalar-Meson Form Factors}
\author{Irina Balakireva}{address={SINP, Moscow State University,
119991 Moscow, Russia}}\author{Wolfgang Lucha}{address={Institute
for High Energy Physics, Austrian Academy of
Sciences,\\Nikolsdorfergasse 18, A-1050 Vienna,
Austria}}\author{Dmitri Melikhov}{address={Institute for High
Energy Physics, Austrian Academy of Sciences,\\Nikolsdorfergasse
18, A-1050 Vienna, Austria},altaddress={Faculty of Physics,
University of Vienna, Boltzmanngasse 5, A-1090 Vienna, Austria}}
\begin{abstract}We scrutinize recent findings on the charged-pion
elastic form factor and the form factors entering in
neutral-meson-to-photons transition amplitudes within the
framework of QCD sum rules.\end{abstract}
\maketitle

\section{Motivation for Rehashing Some Rather Old Stories}
\emph{QCD sum rules\/} relate observable hadronic properties to
the parameters of QCD, the QFT responsible for the formation of
the bound states, by evaluating appropriate correlators of
interpolating currents on the hadron level and on the level of
quarks and gluons, the QCD degrees of freedom. All nonlocal
products of currents may be expressed as series of local operators
by Wilson's \emph{operator product expansion\/}; as a consequence
of this at QCD level correlators obtain perturbative and
nonperturbative contributions, the latter involving the universal
vacuum condensates. \emph{Borel transformations\/} to new
variables---called the Borel parameters---serve to suppress
impacts of both excitations and continuum, and to remove existing
subtraction terms. By rephrasing the perturbative parts of
QCD-level correlators as dispersion integrals over spectral
densities, our ignorance about higher hadronic states can be
hidden by invoking the concept of \emph{quark--hadron duality\/}:
Beyond certain effective thresholds, all the perturbative QCD
contributions are \emph{assumed\/} to cancel those of hadron
excitations and continuum. For \emph{infinitely large\/} Borel
(mass) parameters, all contributions of nonperturbative QCD vanish
and one ends up with QCD sum rules in the limit of \emph{local
duality\/} (LD); these LD sum rules constitute famous tools for
the analysis of form factors. We adopt this approach to revise
anew \cite{blm2011,lm2011} recent dubious findings for
the~charged-pion elastic form factor and the form factor governing
the neutral-pion--$\gamma$ transition $\pi^0\to\gamma\,\gamma^*.$

\section{Dispersive QCD Sum Rules in Local-Duality Limit \cite{ld}}
The dependence of both form factors $F(Q^2)$ on the involved
momentum transfer squared $Q^2\ge0$ may be extracted from two LD
sum rules satisfied by three-current \emph{correlators\/}, of one
vector and two axialvector currents for the charged-pion's
\emph{elastic form factor\/} $F_\pi(Q^2)$ or of one axialvector
and two vector currents for the neutral-pion's \emph{transition
form factor\/} $F_{\pi\gamma}(Q^2),$ involving exclusively
perturbative spectral densities $\Delta(s_1,s_2,Q^2)$ and
$\sigma(s,Q^2),$ respectively, as well as the (weak) decay
constant $f_\pi$ of the charged pion,~$f_\pi=130\;\mbox{MeV}$:
$$F_\pi(Q^2)=\frac{1}{f_\pi^2}\int_0^{s_{\rm eff}(Q^2)}
\hspace{-5.75ex}{\rm d}s_1\hspace{2ex}\int_0^{s_{\rm
eff}(Q^2)}\hspace{-5.75ex}{\rm d}s_2\,\Delta(s_1,s_2,Q^2)\ ,\qquad
F_{\pi\gamma}(Q^2)=\frac{1}{f_\pi}\int_0^{\bar s_{\rm
eff}(Q^2)}\hspace{-5.75ex}{\rm d}s\,\sigma(s,Q^2)\ .$$Any
nonperturbative dynamics is encoded in the effective thresholds
$s_{\rm eff}(Q^2)$ or $\bar s_{\rm eff}(Q^2).$ As power series in
the strong coupling $\alpha_{\rm s},$ the spectral densities are
known up to $O(\alpha_{\rm s})$ or two-loop accuracy
\cite{speden}. Factorization theorems for hard form factors
entail, as \emph{asymptotic\/} form-factor behaviour,
$Q^2\,F_\pi(Q^2)\to8\,\pi\,\alpha_{\rm s}(Q^2)\,f_\pi^2$ and
$Q^2\,F_{\pi\gamma}(Q^2)\to\sqrt{2}\,f_\pi$ for $Q^2\to\infty$
\cite{brodsky}. This feature is reproduced by the LD sum rules if
the effective thresholds behave~like
$$\lim_{Q^2\to\infty}\hspace{-.7ex}s_{\rm
eff}(Q^2)=\lim_{Q^2\to\infty}\hspace{-.7ex}\bar s_{\rm
eff}(Q^2)={4\,\pi^2\,f_\pi^2}\approx0.671\;\mbox{GeV}^2\ .$$The
formulation of reliable criteria for fixing effective thresholds
is highly nontrivial \cite{lms1}: $s_{\rm eff}(Q^2)$ and $\bar
s_{\rm eff}(Q^2)$ won't be equal neither to these asymptotes nor
to each other at finite $Q^2$ \cite{lms2}. In the simplest LD
model \cite{ld} they are approximated at moderate but not too
small $Q^2$ by their asymptotes: $s_{\rm eff}(Q^2)=\bar s_{\rm
eff}(Q^2)={4\,\pi^2\,f_\pi^2}.$ For clarification and
quantification of our concerns, let's introduce the notion of an
\emph{equivalent effective threshold\/}, defined by requiring that
one's sum rule for a form factor, equipped with \emph{this\/}
quantity as its effective threshold, reproduces the experimental
data or a particular theoretical prediction exactly. The accuracy
of any sum-rule approach can be estimated from quantum mechanics:
there all form factors can be found exactly by numerical solution
\cite{Lucha98} of Schr\"odinger equations.

\section{Charged-Pion Elastic Form Factor \cite{blm2011}}
\begin{figure}[b]
\includegraphics[height=.209\textheight,scale=1]{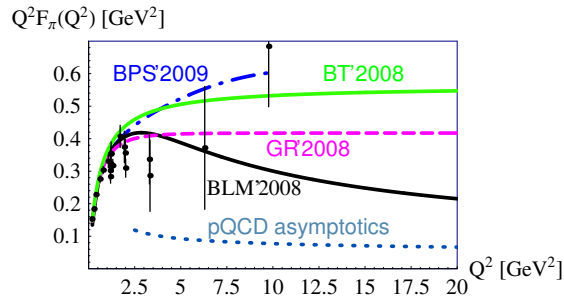}
\caption{Elastic $\pi^\pm$ form factor $F_\pi(Q^2)$:
\emph{selected interpretations\/} \cite{blm2011,recent} of
experimental findings~\cite{data_piff}.}\label{Fig:2}\end{figure}

\begin{figure}[h]
\includegraphics[height=.212\textheight,scale=1]{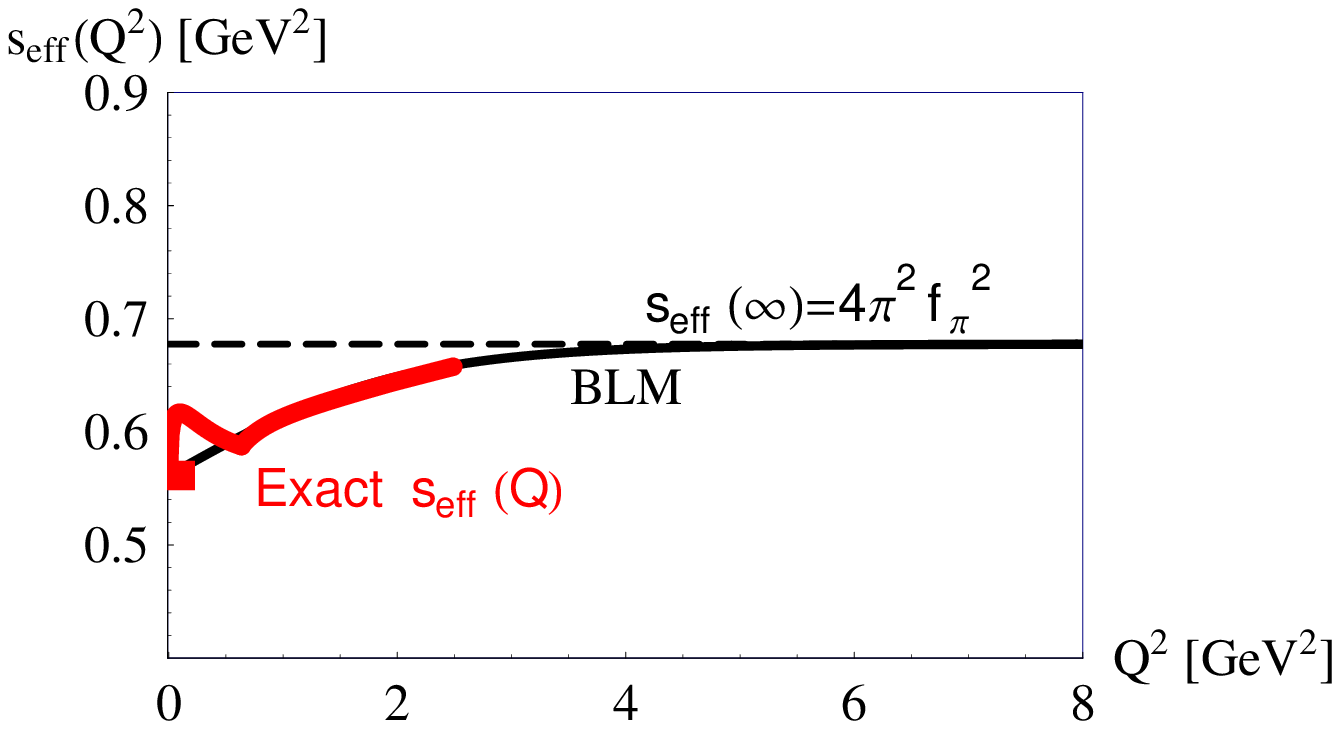}\hspace{-2ex}
\includegraphics[height=.212\textheight,scale=1]{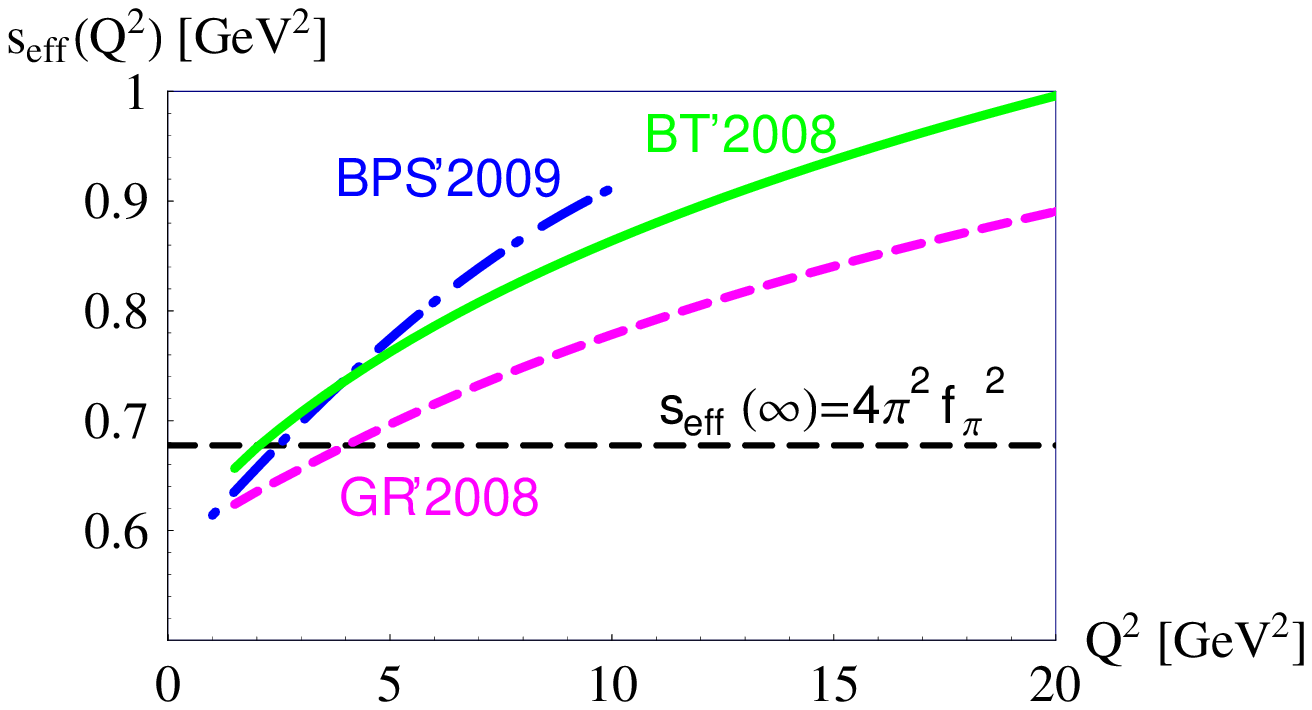}
\caption{Effective threshold $s_{\rm eff}(Q^2)$: most simple
\cite{ld} and somewhat improved \cite{blm2011} LD approaches fit
to its observed \cite{data_piff} `exact' behaviour (left panel)
but are in conflict with \emph{some\/} predictions \cite{recent}
(right~panel).}\label{Fig:3}\end{figure}

Although
the pion belongs to the best-studied meson states, \emph{some\/}
of its properties are not sufficiently well understood. The
\emph{qualitative\/} behaviour of its elastic form
factor,~$F_\pi(Q^2),$ for momentum transfers squared
$Q^2\approx5$--$50\;\mbox{GeV}^2$ gives rise to, or triggers, a
controversy between theory and experiment \cite{data_piff}
(Fig.~\ref{Fig:2}).
Inspecting the present status by adopting our equivalent effective
thresholds, we find that our exact effective threshold, calculated
back from the available experimental data \cite{data_piff}
approaches our (marginally more sophisticated) parametrization
\cite{blm2011} of the effective threshold $s_{\rm eff}(Q^2),$
interpolating between its LD limit and its value at $Q^2=0,$ given
by normalization, at rather low $Q^2$ (Fig.~\ref{Fig:3}, left). In
contrast, recent theoretical analyses \cite{recent} apparently
miss local duality up to $Q^2\approx20\;\mbox{GeV}^2$
(Fig.~\ref{Fig:3}, right). So we conclude that, in the region
$Q^2=20$--$50\;\mbox{GeV}^2$, sizable deviations of $F_\pi(Q^2)$
from the LD expectations, as predicted by some analyses
\cite{recent}, seem to be rather unlikely.

\section{Form Factors for $\eta,$ $\eta^\prime$ Meson Transitions
$\eta^{(\prime)}\to\gamma\,\gamma^*$ \cite{lm2011}}
\begin{figure}[h]
\includegraphics[height=.213\textheight,scale=1]{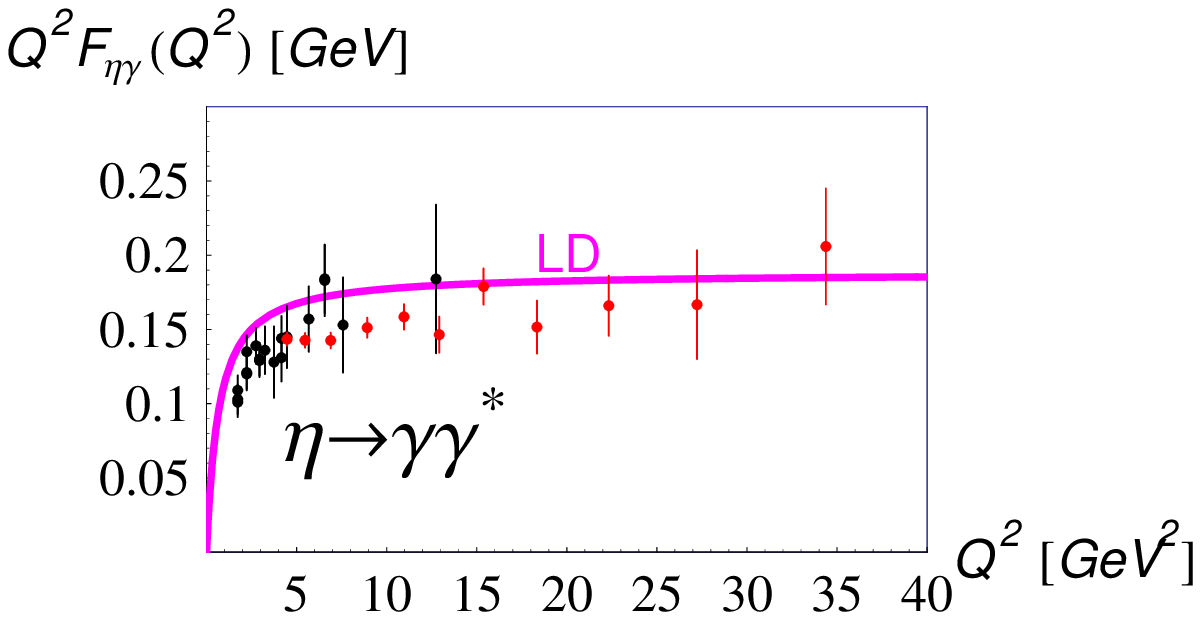}\hspace{-2ex}
\includegraphics[height=.213\textheight,scale=1]{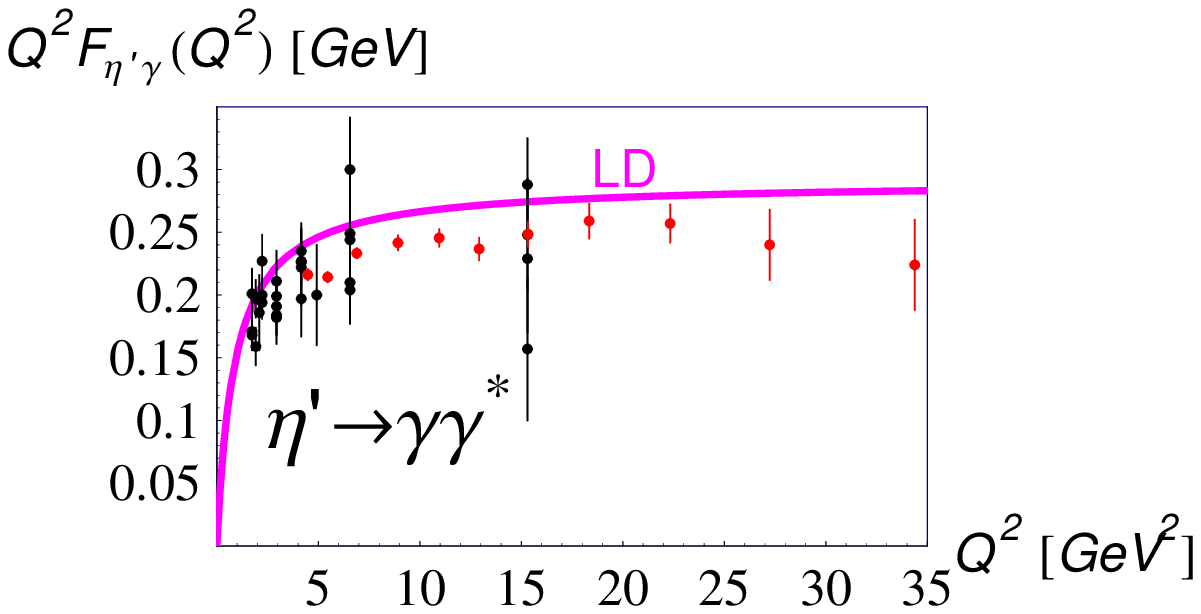}
\caption{Transition form factors
$F_{(\eta,\eta^\prime)\gamma}(Q^2)$: Local duality and observation
\cite{cello-cleo,babar1} agree perfectly.}\label{Fig:4}
\end{figure}

The
\emph{flavour structure\/} of the isoscalar mesons $\eta,$
$\eta^\prime$ is a linear combination of $\bar uu,$ $\bar dd,$ and
$\bar ss.$ The non-ideal mixing of $\eta,$ $\eta^\prime$ is
reflected by their transition form factors
$F_{(\eta,\eta^\prime)\gamma}(Q^2)$ receiving non-strange and
strange contributions. We obtain for the transition form factors
of both $\eta$ and $\eta^\prime$ the anticipated
\emph{agreement\/} between LD and experiment
\cite{cello-cleo,babar1} (Fig.~\ref{Fig:4}).

\section{Form Factor for Neutral-Pion Transition
$\pi^0\to\gamma\,\gamma^*$ \cite{lm2011}}
\begin{figure}[ht]
\includegraphics[height=.212\textheight,scale=1]{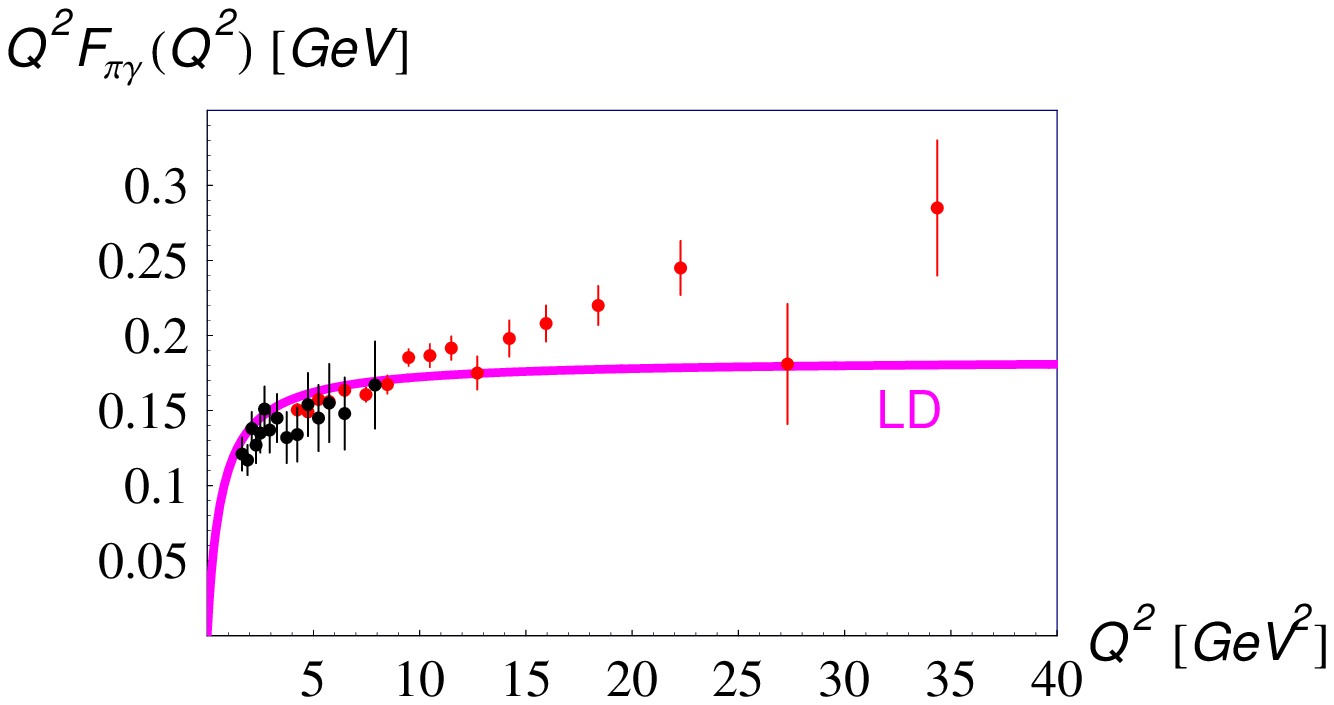}\hspace{-2ex}
\includegraphics[height=.212\textheight,scale=1]{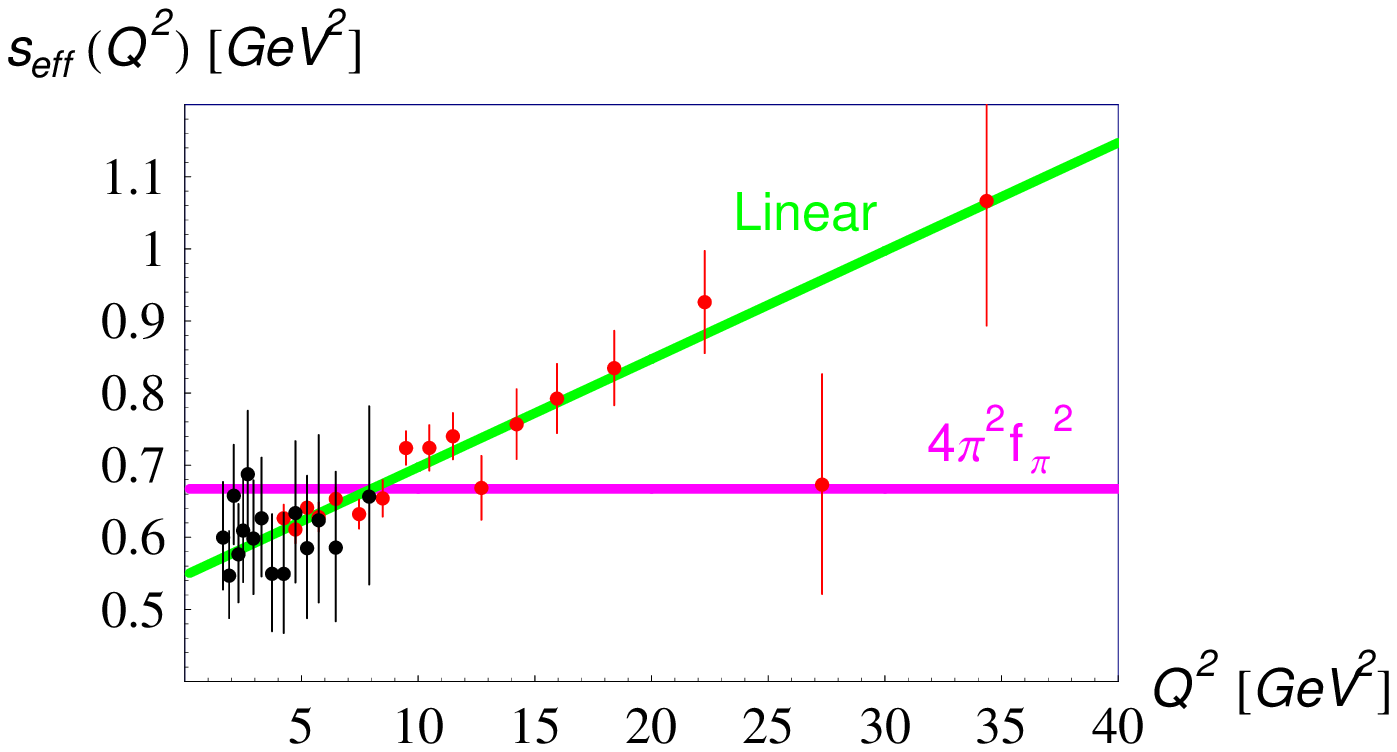}
\caption{Form factor $F_{\pi\gamma}(Q^2)$ for the $\pi^0$
transition $\pi^0\to\gamma\,\gamma^*$: the theoretically
unexpected and rather surprising deviation of the {\sc BaBar} data
\cite{babar} from the expectations of LD (left panel) causes our
equivalent effective threshold $\bar s_{\rm eff}(Q^2)$ to rise
linearly without any sign of caring about the LD prediction~(right
panel).}\label{Fig:5}\end{figure}

Interestingly, some of
the measurements \cite{cello-cleo,babar,Belle} of the neutral
pion's transition form factor $F_{\pi\gamma}(Q^2)$ evince a rapid
growth with $Q^2$ that still awaits an explanation
(Fig.~\ref{Fig:5}, left). However, a recent Belle measurement
\cite{Belle} definitely contradicts the {\sc BaBar} findings for
$F_{\pi\gamma}(Q^2).$ We conclude that the {\sc BaBar} results
must be taken with due care; see also~\cite{findings}.

\begin{theacknowledgments}D.M.\ has been supported by the Austrian
Science Fund (FWF) under project
no.~P22843.\end{theacknowledgments}

\bibliographystyle{aipproc}\end{document}